\documentclass[
aip,
jap,
numerical,
reprint,
amsmath,
amssymb]{revtex4-1}
\usepackage{graphicx}
\usepackage{dcolumn}
\usepackage{bm}
\usepackage{textcomp}
\usepackage{revsymb}
\usepackage{amssymb}
\usepackage{ifsym}
\usepackage{wasysym}
\usepackage{txfonts}
\usepackage{longtable}
\usepackage{tipa}

\begin{document}




\title{KSbOSiO$_4$ microcrystallites as a source of corrosion of blue-green lead-potassium glass beads of the 19th century}

\author{T.~V.~Yuryeva}
\email{tvyur@kapella.gpi.ru} 
\affiliation{The State Research Institute for Restoration of the Ministry of Culture of Russian Federation, bldg 1, 44 Gastello Street,  Moscow, 107114, Russia}

\author{I.~B.~Afanasyev}
\affiliation{Forensic Science Center of the Ministry of the Interior of the Russian Federation, 5 Zoya and Alexander Kosmodemyansky Street, Moscow, 125130, Russia} 

\author{E.~A.~Morozova}
\altaffiliation[Also at ]{Kurnakov Institute of General and Inorganic 
Chemistry, Russian Academy of Sciences, Moscow, Russia}
\affiliation{The State Research Institute for Restoration of the Ministry of Culture of Russian Federation, bldg 1, 44 Gastello Street,  Moscow, 107114, Russia}

\author{I.~F.~Kadikova}
\affiliation{The State Research Institute for Restoration of the Ministry of Culture of Russian Federation, bldg 1, 44 Gastello Street,  Moscow, 107114, Russia}

\author{V.~S.~Popov}
\affiliation{Kurnakov Institute of General and Inorganic Chemistry of the Russian Academy of Sciences, 31 Leninsky Avenue, Moscow, 119071, Russia}

\author{V.~A.~Yuryev}
\homepage[Home page: ]{http://www.gpi.ru/eng/staff\_s.php?eng=1\&id=125}
\email{vyuryev@kapella.gpi.ru} 
\affiliation{A.\,M.\,Prokhorov General Physics Institute of the Russian Academy of Sciences, 38 Vavilov Street, Moscow, 119991, Russia}

\date{\today}%

\begin{abstract}
Presently, deterioration of glass beads is a significant problem in conservation and restoration of beaded exhibits in museums. 
Glass corrosion affects nearly all kinds of beads but cloudy blue-green ones are more than others subjected to disastrous destruction.
However, physical and chemical mechanisms of this phenomenon have not been understood thus far.
This article presents results of a study of elemental and phase composition of glass of the blue-green beads of the 19th century obtained from exhibits kept in Russian museums. 
Using scanning electron microscopy, X-ray microanalysis and X-ray powder analysis
we have detected and investigated Sb-rich microinclusions in the glass matrix of these beads and found them to be micro crystallites of KSbSiO$_5$.
These crystallites were not detected in other kinds of beads which are much less subjected to corrosion than the blue-green ones and deteriorate in a different way.
We believe that individual precipitates of KSbSiO$_5$ and especially their clusters play a major role in the blue-green bead deterioration giving rise to slow internal corrosion of the bead glass. 
 
\end{abstract}



\maketitle


\section{\label{intro}Introduction }

Beadwork in the works of arts and crafts has existed as a substantive art from  the late  18th century till the 1880th.\footnote{It is known from archeology and history of arts, however, that beads have been used 	for adornment, especially for ornamentation of clothes or decorative weaving, during the whole human history since the very ancient time.\cite{History_of_Beads}
}$^,$\cite{History_of_Beads}
At that time, bead embroidery was very popular throughout Europe.\cite{Beads-19th_century_French,*Beads-19th_century}  
That is why almost all art or history museums of the world have collection of beadworks  among  works of arts and crafts. 
In Russia, even in small local history museums,  several exhibits with bead embroidery are always present among household items of that time. 

\begin{figure*}[t]
\includegraphics[scale=1.1]{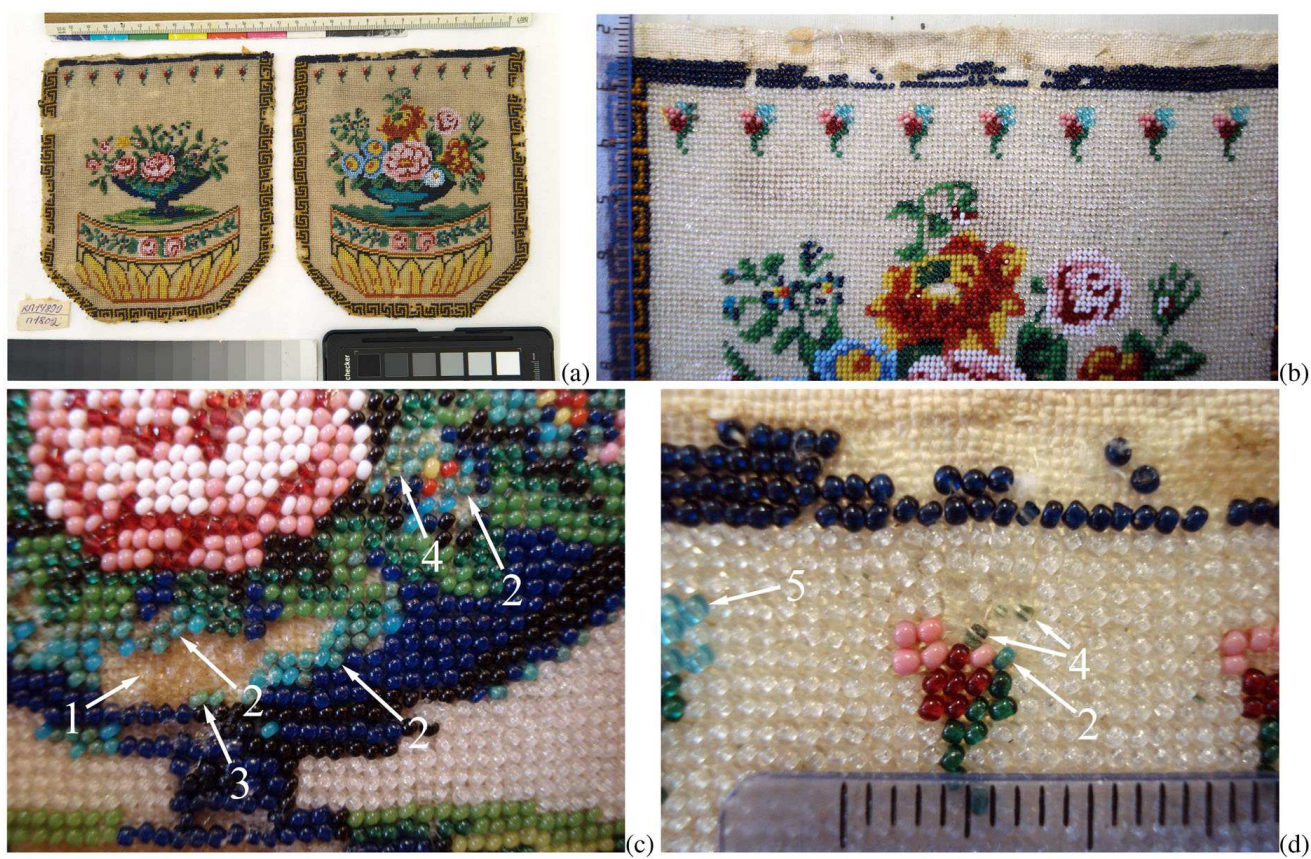}
\caption{\label{fig:beadwork} (Color online)
Fragments of a beaded box of the 19th century, collection of the Museum of A.~S.~Pushkin in Moscow, 
(a) 
before restoration and
(b)
in the process of restoration; 
numerous damages of the beaded picture
are seen in panel (a); 
destructed historic blue-green beads are almost entirely replaced by larger modern blue transparent beads on the top of the beadwork in panel (b).
(c,\,d) 
Photographs of damaged parts of the beaded box; numbers show examples of damages of  the beadwork connected with the deterioration of blue-green beads: 
1 denotes a void appeared  as a result of total destruction of turquoise beads, 
2, 
3 
and 
4 indicate some of blue-green beads at different phases of deterioration shown in 
Fig.\,\ref{fig:destruction}; 
number 2 corresponds to Fig.\,\ref{fig:destruction}\,d, 3 corresponds to Fig.\,\ref{fig:destruction}\,f  and 4 corresponds  to  Fig.\,\ref{fig:destruction}\,g; number 5 shows modern beads replacing historical ones; the scale minor ticks are given in millimeters.
}
\end{figure*}

As a rule, 20 to 30 varieties of glass beads were used for embroidering a beadwork; the total number of varieties, which differ in size, color and type (round, faceted, multi-layered, transparent, cloudy, opaque, etc.), was more than a thousand.\cite{Yurova}
When used in household beaded items, beads were affected by sebum, moisture, detergents, mechanical impacts, etc.; as a result, they eventually deteriorated. 
But even under the careful museum keeping of beaded articles,  beads that have already lived a life in objects used in everyday life or in religious purposes, continue to deteriorate and crumble (Fig.\,\ref{fig:beadwork}). 
This applies to all kinds of beads of different colors  but the cloudy blue-green (or turquoise)  beads are  exposed to the destruction especially strongly. Usually they are referred to as ``unstable beads'' because of their disastrous deterioration. 
Corroded museum blue-green beads of the first half of the 19th century resemble in the character of their destruction  beads of the same color found at archaeological sites.\cite{Arch-Mos-Reg} 
Now one can hardly find an artifact of beads which would not be damaged by corrosion. 
The problem is so acute that  most of the exhibits made of the glass beads may be lost in the near future or original beads would be replaced by modern ones, differing both in the size and colors, in the process of extremely time-consuming and expensive restoration.
That is why the study of causes of degradation of the blue-green glass beads is 
strongly required for the conservation of authentic historic beadworks in 
museum collections.\cite{Yuryev_JOPT}

\begin{figure*}[th]
\includegraphics[scale=1.1]{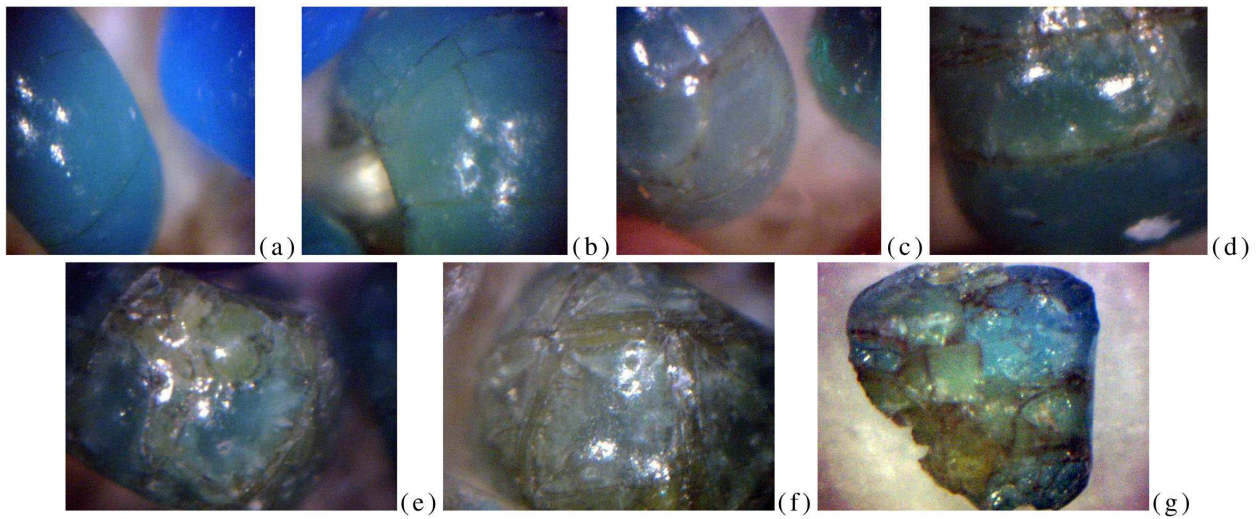}
\caption{\label{fig:destruction} (Color online)
Micro photographs of the blue-green beads illustrating consecutive phases of the glass corrosion: 
(a) cracks appear in the blue beads; 
(b) cracking increases, the color starts to change into
the greenish tone; 
(c),\,(d) beads undergo further changes of the color, cracks change their color to the greatest degree and become brown-green; 
(e) discoloration of beads begins, the surface corrosion develops; 
(f) beads become dim and grainy; 
(g) beads fragment.\cite{Yuryev_JOPT}
}
\end{figure*}

\begin{figure*}[t]
\includegraphics[scale=.98]{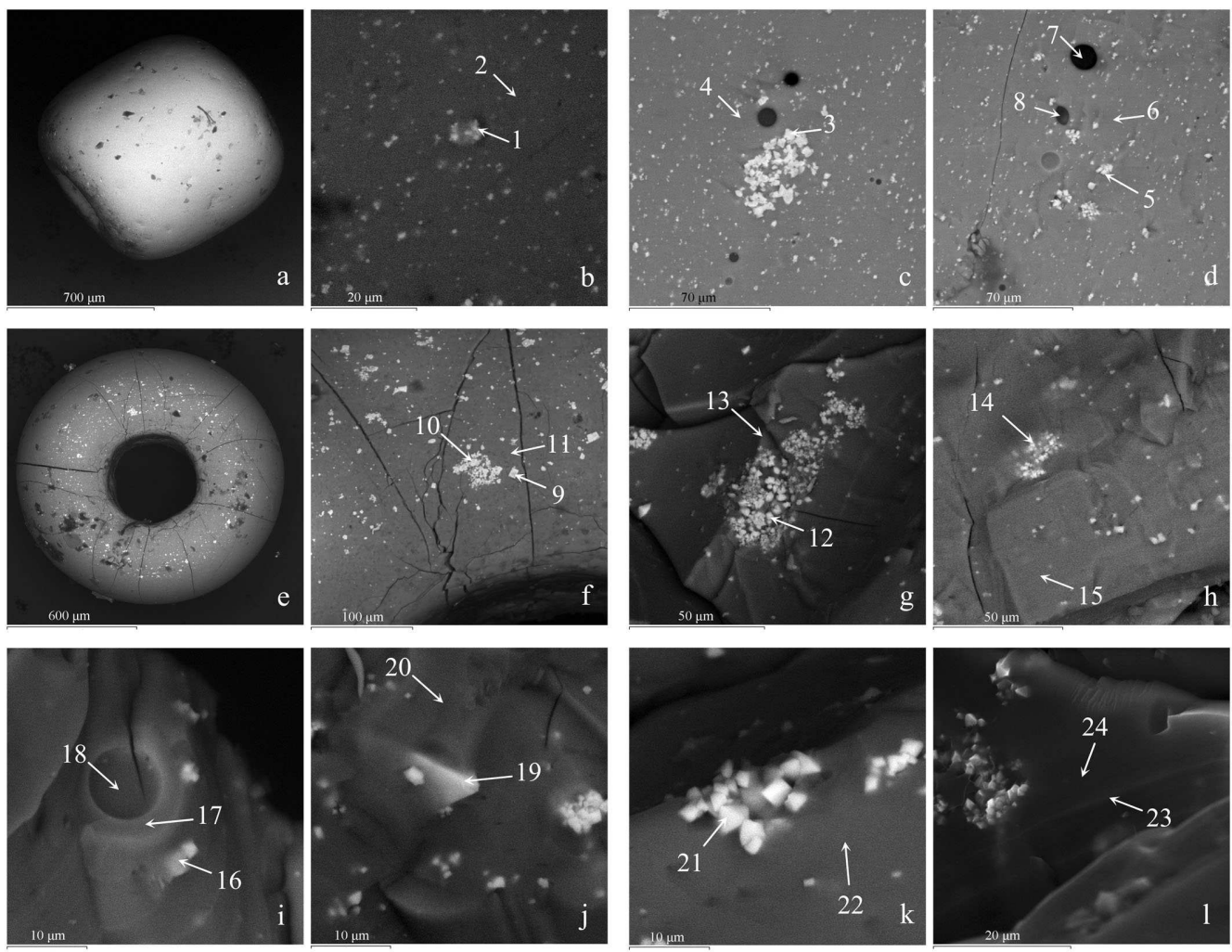}
\caption{\label{fig:SEM}
SEM images (BSE) of blue-green beads at different phases of glass corrosion: (a, b) an intact blue bead, (c, d) a cleft blue bead (on the cleavage surface), (e, f) a strongly degraded bead with blue and brown-green (heavily cracked) segments, (g-l) broken blue-green beads (on the cleavage surface); the scale is 700 {\textmu}m in panel (a), 20 {\textmu}m in panels (b) and (l), 70 {\textmu}m in panels (c) and (d), 600 {\textmu}m in panel (e), 100 {\textmu}m in panel (f), 50 {\textmu}m in panels (g) and (h), 10 {\textmu}m in panels (i), (j) and (k); elemental composition of the substance at the indicated points is presented in Table\,\ref{tab:EDS}.
}
\end{figure*}

The majority of data on this issue existing presently in the scientific 
literature have been obtained for earlier archeological or ethnographic beads 
(see, e.g., Refs.\,\onlinecite{Deteriorating_Glass_Beads, 
MOHAWK_GLASS_TRADEBEADS, *Datch_TRADEBEADS, EuroBeads, 
Canadian_beads_Deterioration, Book_Adam_Lovell, Etn_Conc_Newsletter, 
*BEAD_FORUM, *Postprints_journal, Archaeometrical_analysis, 
Wood_African_beads1, *Wood_African_beads, Prinsloo_Raman_African_1, 
*Prinsloo_Raman_African_2}).\footnote{
	It should be noted that beads for embroidery and beads for manufacture of 
	necklaces, rosaries, chaplets, bangles, etc. often are not distinguished 
	between in the literature despite the difference in their sizes as well as 
	manufacturing processes.
} 
It is believed that glass bead deterioration processes are caused only by the 
chemical reactions occurring on the surface, and only they, by analogy with a 
stained glass and a dish glass, are responsible for the bead corrosion.
The process of deterioration of the glass beads have usually been associated by researchers with the impact of external factors on the glass, especially atmospheric moisture and\,/\,or a cloth or a leather on which the beads were sewn,\cite{Etn_Conc_Newsletter, *BEAD_FORUM, *Postprints_journal}
resulting in so called crizzling,\cite{Crizzling_Problem_1975} a specific roughening or cracking of the glass surface.\footnote{
Etymology of the term `crizzling' is connected with the appearance of the rough and cracked glass surface resembling sunburnt skin on face or hands.\cite{Crizzling_Problem_1975}}$^,$\footnote{
See also Ref.\,\onlinecite{Glass_corrosion_multiscale, *Stained_glass_window} for	effect of atmospheric water on corrosion of potassium glass
and a series of articles presented in 
Ref.\,\onlinecite{Robinet_1, *Robinet_4, *Robinet_2, *Robinet_3} 
for effect of organic polutants including acid from wooden cabinets or showcases on glassware corrosion}$^,$\cite{Glass_corrosion_multiscale, *Stained_glass_window}$^,$\cite{Robinet_1, *Robinet_4, *Robinet_2, *Robinet_3}
Such corrosion flows in several stages: the moisture comes into contact with the surface of the glass, the glass leaches, forms a near-surface layer depleted in alkali metals, this layer cracks, moisture penetrates deeper, cracking process increases, and the glass breaks into fragments.
However, this scenario contradicts the fact that similar beads at different phases of deterioration---intact, slightly cracked, severely cracked and changed the color or discolored, partially or completely fragmented---are often present on a single exhibit in immediate neighbourhood (Figs.\,\ref{fig:beadwork} and \ref{fig:destruction}): they have obviously  been subjected to the same impact of moisture but corroded to different degree. 
Thus, it is necessary to identify the real mechanisms of deterioration of the historic blue-green beads.

The authors of this article have recently proposed an alternative mechanism for the destruction of the blue-green beads associated with the physical and chemical processes in the bulk of the glass mass.\cite{Yuryev_JOPT}
Bead degradation may occur due to internal stress, resulting in local inelastic strain and rupture of glass.
It is assumed that some sufficiently large precipitates, probably crystalline, may be sources of the internal stress.
Apparently, they are present in the  glass of beads since microimages of some inclusions resemble typical images of faceted crystallites which may arise as a result of the decomposition, the diffusion and the crystallization of some chemical  components of the glass, such as pigments, opacifiers, fluxes or stabilizers, in the manufacturing process of the glass beads.
Arising, they generate a stress field that causes internal cracking of the glass in the process of cooling and additional gettering of impurities and chemical components dissolved in the glass, which, in turn, form a new generation of micro precipitates and impurity atmospheres around the micro cracks and further contribute to internal corrosion of beads. 
These processes can be extremely long-lasting since after cooling of beads they occur at room temperature at which both the diffusion rates of components and the rates of chemical reactions as well as the formation rates of the precipitates are low. 
However this hypothesis requires careful experimental verification.
If the proposed hypothesis is correct the process of destruction of the blue-green beads was already embedded in its production cycle.
In this case, the long-term forecast for the safety of this type of bead is unfavorable and no chemical conservation methods allow one to save it;  exhibits of this kind of beads likely can be saved if kept at a lowered temperature.

Preliminary studies conducted by the authors of the article show that the processes of deterioration of other types of beads, e.g. the red ones, which are also considered as unstable,\cite{Etn_Conc_Newsletter, *BEAD_FORUM, *Postprints_journal} are different from those of the blue-green ones which also contradicts the theory of chemical surface corrosion. 

So, corrosion mechanisms of the blue-green bead are currently not understood.

This article presents experimental results on composition and structure of the  secular blue-green beads of the 19th century. We have demonstrated that the previously observed micro inclusions are crystalline. We have investigated the composition and crystal system of these precipitates using the scanning electron microscopy, the X-ray microanalysis and the X-ray phase analysis and found them to be crystallites of 
KSbSiO$_5$.
Other explored beads of different kinds do not contain such precipitates and even if they contain Sb-rich inclusions the latter do not manifest any attributes of crystallinity.

\begin{figure}[th]
\includegraphics[scale=1]{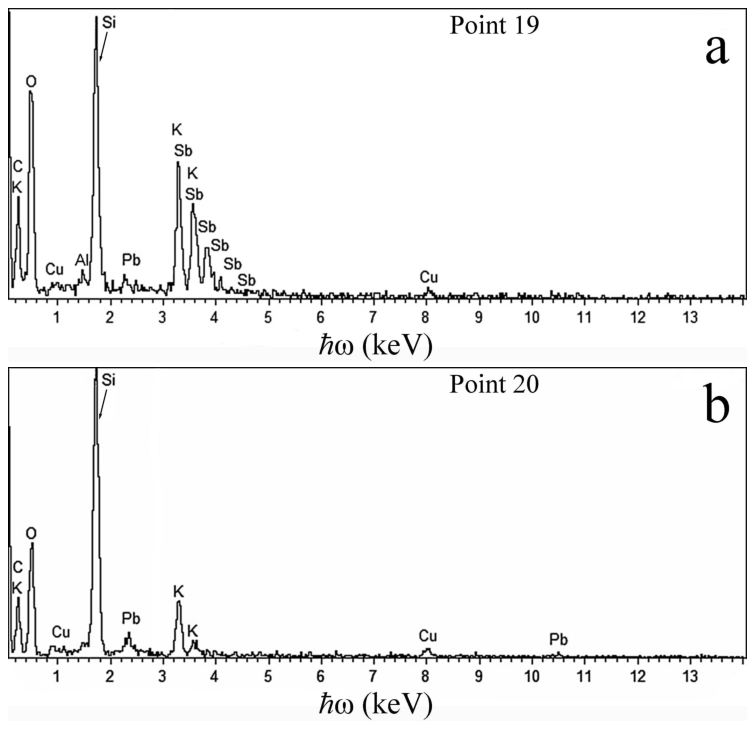}
\caption{\label{fig:EDX}
Examples of EDS spectra obtained at different points on samples of the blue-green beads (Fig.\,\ref{fig:SEM}, Table\,\ref{tab:EDS}); panel (a)  demonstrates the elemental composition of the precipitates and glass around them within the electron probe area, panel (b) illustrates the composition of the glass bulk free of the precipitates.
}
\end{figure}


\section{Samples and Methods}

Experimental samples of color glass beads were obtained during restoration of historic beaded articles of the 19th century from museum collections (an example is shown in  Fig.\,\ref{fig:beadwork}).
The main attention was paid upon the cloudy blue-green beads at different phases of deterioration (Fig.\,\ref{fig:destruction}, Table\,\ref{tab:EDS}).\cite{Yuryev_JOPT, SEM-15} 
They were examined by means of the scanning electron microscopy (SEM), X-ray microanalysis and X-ray diffractometry.
Beads of different colors (yellow, white and red) were also investigated using SEM but only yellow ones were analysed using the X-ray diffraction.
Before the experiments, the samples were washed with ethanol at 40{\textcelsius} for 20 minutes in a chemical glass placed into an ultrasonic bath (120 W, 40 kHz).


For the SEM studies, Tescan Vega-II XMU Scanning electron microscope was used in the mode of backscattered electrons (BSE). The X-ray microanalysis was carried out using Inca Energy 450 energy dispersion spectrometer (EDS). 


The X-ray phase analysis was performed by means of the Debye-Scherrer powder diffraction method\cite{Debye-X-ray_diffraction_article, X-ray_Diffraction_Book} using the Bruker~D8 Advance diffractometer at non-monochromatic Cu\,K$_{\alpha}$  band (Cu\,K$_{\alpha_{1,2}}$, $\lambdaup$ = 1.542 \r{A}); 
the diffraction patterns were scanned in a $2\thetaup$ interval from 5 to 75$^{\circ}$ with the  steps of $0.02^{\circ}$; the data acquisition time per one point was 9.0 sec. 
PDF-2 Powder Diffraction Database (Joint Committee on Powder Diffraction Standards\,--\,International Centre for Diffraction Data) was used for the phase composition analysis. In addition, Crystallography Open Database (COD) was also used for the data analysis.\cite{COD_American_Presentation, *COD_presentation-2, *COD_presentation-1, *COD_error_correcting}

\begin{figure}[th]
\includegraphics[scale=.5]{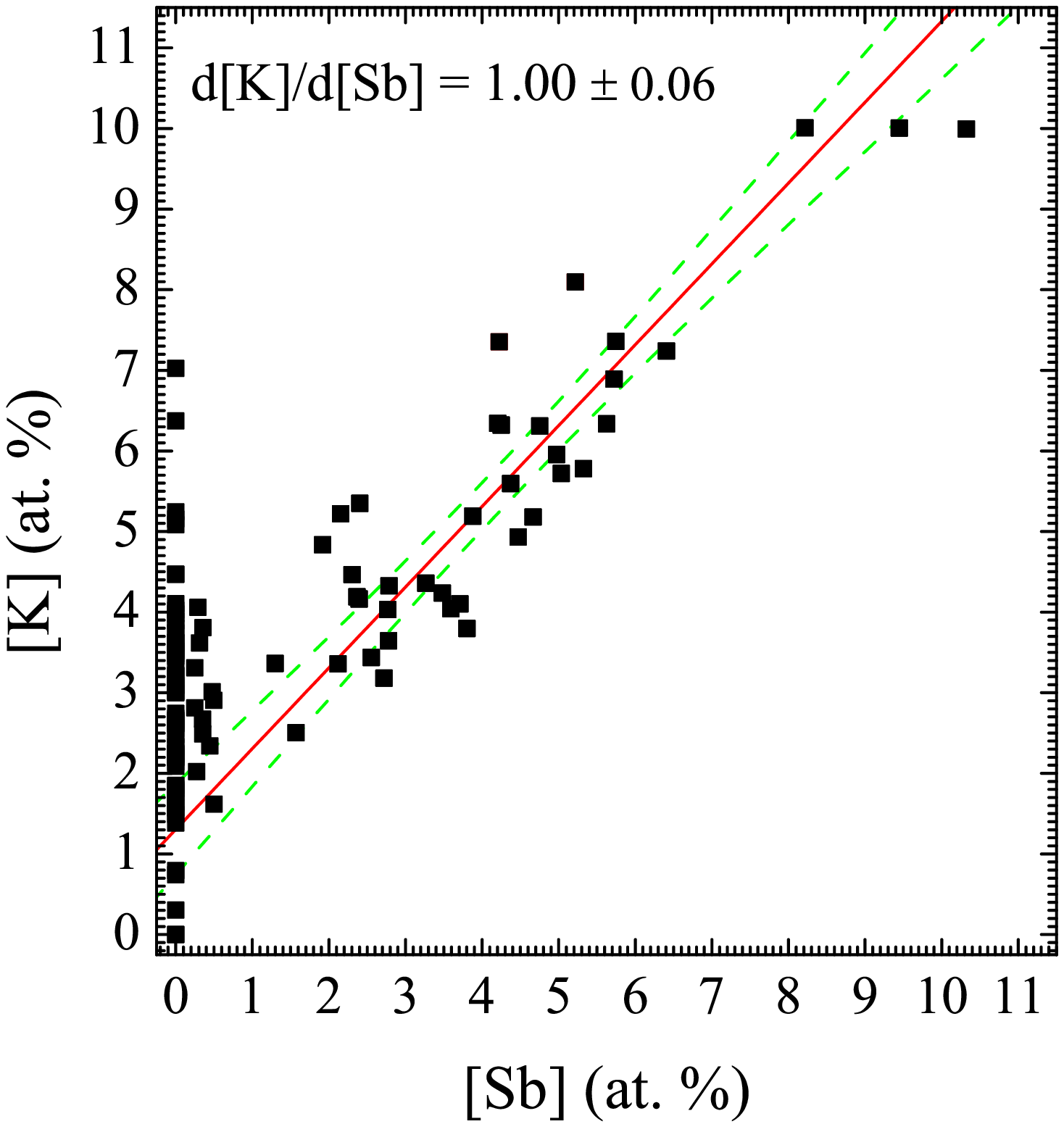}
\caption{\label{fig:K_vs_Sb}(Color online)
A dependence of the potassium concentration [K] on the antimony concentration [Sb] determined from the  data of EDS obtained at different points of analysis on blue-green beads. The best-fit straight line is plotted for [Sb] $>$ 1 at.\,\%, i.e. for regions of the Sb-rich precipitates within the probed areas, the line slope is $1.00\pm 0.06$; the dashed lines show 95-\% confidence bands. [K] values for [Sb] $<$ 1 at.\,\% reflect the potassium content in the glass bulk at different points of the examined samples.}
\end{figure}

\begin{figure*}[th]
\includegraphics[scale=1]{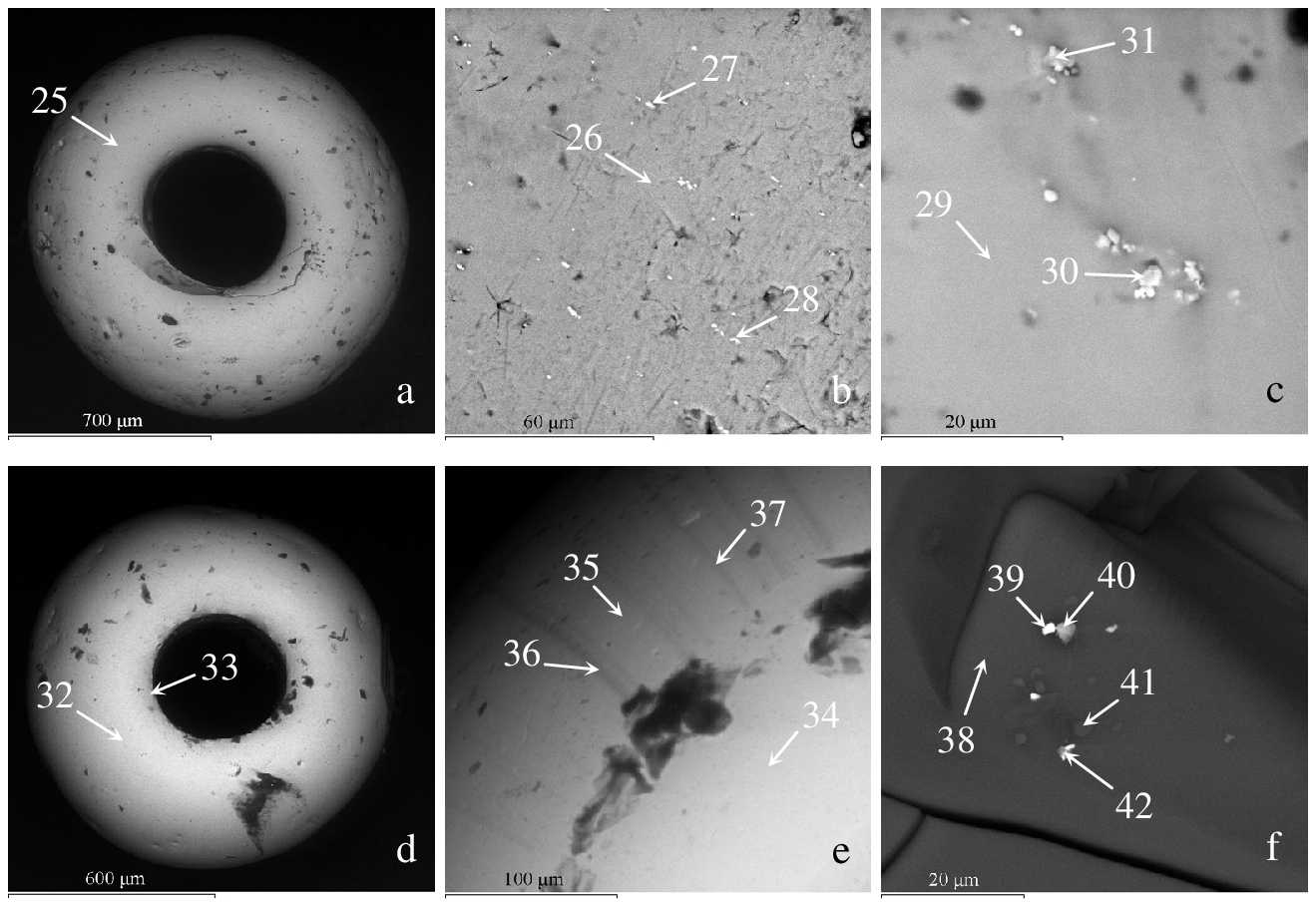}
\caption{\label{fig:SEM-1}
SEM images (BSE) of yellow [intact (a) and cleft (b, c)], white [intact, opaque (d) and less opaque (e)] and cleft faceted red  (f) beads; the scale is 700 {\textmu}m in panel (a), 60 {\textmu}m in panel (b), 20 {\textmu}m in panels (c) and (f), 600 {\textmu}m in panel (d), and 100 {\textmu}m in panel (e); elemental composition of the substance at the indicated points is presented in Table\,\ref{tab:EDS}.
}
\end{figure*}

\section{Microscopy and analysis}


\subsection{Scanning electron microscopy and X-ray microanalysis}

SEM micrographs of the ``unstable'' blue-green beads (intact, cleft, strongly degraded and broken) are presented in Fig.\,\ref{fig:SEM}. Numerous faceted precipitates and their clusters are observed both on the bead surfaces and in the glass bulk in all the beads independently of the stage of the deterioration. We have analyzed the elemental composition of the crystallites and the glass at the points indicated by arrows in the panels of Fig.\,\ref{fig:SEM} and found the following (Table\,\ref{tab:EDS}). (i) The blue-green beads are made of lead-potassium glass sometimes containing a considerable amount of aluminum and always doped by copper. (ii) A considerable content of potassium is observed in all points of the analysis independently of whether the crystalline inclusion or the glass bulk is analyzed;  calcium is also detected at some points. (iii) Antimony is observed only in the crystallites; it has never been detected in the glass bulk (Fig.\,\ref{fig:EDX}).\footnote{
Note that the only exception was detected at point 23; in this sample striation was observed with the stria containing some amount of Sb (Fig.\,\ref{fig:SEM}, panel l; Table\,\ref{tab:EDS}).
}
(iv) Other elements are either absent in the samples or present in trace amounts.

We have analysed correlations between contents of atoms of the detected elements measured at the same point and found the only correlation between potassium and antimony (Fig.\,\ref{fig:K_vs_Sb}):\footnote{
A wider data array than that presented in Table\,\ref{tab:EDS} was employed in the correlation analysis.
} 
for the antimony content [Sb] $>$ 1 at.\,\%, i.e. for regions where the Sb-rich precipitates occupy a large enough fraction of the probed areas, the line slope is equal to $1.00\pm 0.06$. [K] values for [Sb] $<$ 1 at.\,\% reflect the potassium content in the glass bulk at different points of the examined samples. This result makes us conclude that K and Sb are contained in the crystallites in the same proportion and no other elements except for  silicon and oxigen can compose the precipitates;\cite{SEM-16} the compounds satisfying this requirement are KSb, KSbO$_3$ and KSbSiO$_5$. 

Beads of different colors have also been investigated by means of SEM and EDS (Fig.\,\ref{fig:SEM-1}, Table\,\ref{tab:EDS}). Like the blue-green beads, 
the white ones are made of lead-potassium glass; the yellow beads are made of lead glass but the mass fraction of lead is much higher in them reaching 45 wt.\,\%;
some amount of Pb ($\sim 4.5$ wt.\,\%) was also detected in some inclusions in the red beads.
All these beads contain calcium; potassium virtually is not contained only in yellow beads. 
The white beads contain sodium in the glass bulk;
some aluminum is present in the semitransparent white beads.
Antimony was detected only in the yellow beads  but in much less proportion than in the blue-green ones.  

Like in the blue-green beads, antimony is not contained in the glass bulk of the yellow beads but only in inclusions. However, these inclusions are not faceted; potassium is absent in their composition.
They may be composed by some yellow pigments (e.g., Sb$_2$O$_5\cdotp n$PbO which have been known since the ancient times and often used by painters since the Middle Ages). 

The glass bulk of the faceted red  beads contain only calcium and potassium (Fig.\,\ref{fig:SEM-1}\,f, point 38). Microinclusions of various types are observed on cleavages of these beads;  
they contain sets of different chemical elements, e.g., sulfur, arsenic and mercury (point 39), sulfur, zinc and barium (point 40), magnesium, aluminum, iron and lead (point 41), antimony, gold and lead (point 42); some of the particles look faceted.

It should be noted that beads containing less antimony (e.g., yellow or red) or not containing it (white) are significantly less subjected to destruction than the blue-green ones and degrade (if degrade) in a different manner. For instance,
deterioration of the faceted red  beads occurs through the formation of flakes on the facets. The cold faceting during the manufacture presumably left microcracks on the facets of the beads which grew over time. These beads usually cleave on edges between the facets.

\subsection{X-ray phase analysis}

\begin{figure}[ht]
\includegraphics[scale=.55]{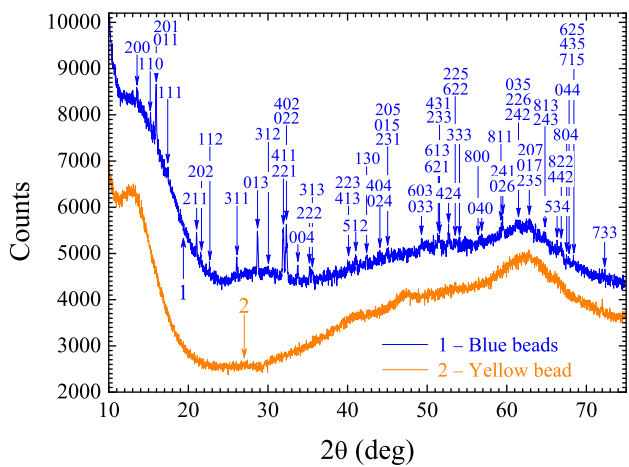}
\caption{\label{fig:XRD}(Color online) 
X-ray powder patterns (Cu\,K$_\alpha$) from samples containing Sb-rich inclusions 
(Figs.\,\ref{fig:SEM} and \ref{fig:SEM-1}, Table\,\ref{tab:EDS}):
(1) a Debye pattern obtained from
a powder of three grinded intact blue beads; the most intense bands identified as those corresponding to KSbOSiO$_4$ are marked with indication of Miller indices of the reflecting
crystal planes (h\,k\,l);
a complete list of reflexes observed in this sample is given in Table\,\ref{tab:diffract}. (2) A powder pattern of  the cleft yellow bead.
 }
\end{figure}

To determine the exact phase composition of the precipitates in the blue-green 
beads, we have investigated the X-ray diffraction by the beads using the 
Debye-Scherrer method. A powder pattern of several grinded intact blue-green 
beads is presented in Fig.\,\ref{fig:XRD} (curve 1). The most of the detected 
bands were identified as those relating to KSbOSiO$_4$ (a complete list of 
reflexes observed in this sample is given in Table\,\ref{tab:diffract}). So, we 
can confidently state that the precipitates in the blue-green beads are 
crystallites of orthorhombic KSbOSiO$_4$.

Some lines have not been identified. They may be connected with a crystalline SiO$_2$ phase in the glass bulk of the beads but this identification is uncertain and rather unreliable; it requires further clarification. 

A diffraction pattern of the cleft yellow bead, which contains antimony-rich inclusions,  is also shown for the comparison in Fig.\,\ref{fig:XRD} (curve 2). This bead is seen to not manifest any signs of  presence of a developed crystalline phase:  intense peaks are not seen in the pattern, except for maybe a line at $2\thetaup \approx 72^{\circ}$, which is also observed in the powder pattern of the blue-green beads,  although some weak lines are present in this  diffraction pattern. This bead certainly does not contain KSbOSiO$_4$.

\section{Discussion}

Presently, we do not realize the origin of the KSbOSiO$_4$ crystals in the beads. 
Antimony might enter the glass accidentally as an uncontrolled impurity from charcoal ash which was often used as a source of potassium or from minerals. 
Alternatively, antimony oxide 
could be intentionally added to the glass charge as an opacifier: antimony was used for opacifying beads until the mid 19th century.\cite{EuroBeads} 

The chemical and physical  processes of KSbOSiO$_4$ precipitation in the glass bulk, either during cooling of the melt or the glass mass or in the course of time as a result of diffusion and solid-state chemical reactions at room temperature, are unknown at present.  
However, we can make assumption about possible conditions of the KSbOSiO$_4$ formation in glass, which, we believe, looks rather grounded. 
Orthorhombic KSbOSiO$_4$ is known to form at the temperature higher than 1100{\textcelsius}.\cite{COD-KSbOSiO4}$^,$\footnote{
A higher temperature  of the phase transition between tetragonal and orthorhombic phases of KSbOSiO$_4$ (1200{\textcelsius}) is given in Ref.\,\onlinecite{Tetragonal-orthorhombic}
}$^,$\cite{Tetragonal-orthorhombic}
At lower temperatures mainly the tetragonal KSbOSiO$_4$  appears. 
Since no tetragonal KSbOSiO$_4$ was observed in our experiments in bead glass we believe that the silicate crystals formed at the temperature higher than 1100{\textcelsius}; most likely, they appeared during  glass melting ($T = 1300$ to 1400{\textcelsius}) and bubbling ($T>1500$\textcelsius).\cite{Glass_Technology}

Let us consider two scenarios of the bead destruction connected with the emergence of  the KSbOSiO$_4$ crystallites.
Once arising, the crystallites should generate strain fields during glass cooling after solidification of melt due to a difference of thermal expansion coefficients of glass and KSbOSiO$_4$ crystals. The emerged stress results in glass rupture partially relieving stain;  micro cracks should develop in the regions of the maximum strain.
Note that the tensile  strain should emerge to keep the bead total volume 
unchanged and to allow cracks to stay in the internal domains of beads since 
otherwise, if compressive strain emerges, cracks would always reach the surface 
to increase the total volume and relieve the strain and beads would fragment 
mainly during cooling or maybe during embroidering.
Over time, the stress should be relieved by increasing the previously formed microcracks and appearing new ones. Finally, glass should become entirely cracked and a bead should fragment. We see numerous individual crystallites and large colonies of the KSbOSiO$_4$ precipitates on surfaces of bead fragments (Fig.\,\ref{fig:SEM}). 

As an alternative, the KSbOSiO$_4$ crystallites may stimulate glass crystallization resulting in changes in volume of the crystallized domains that, in turn, results in glass rupture and granulation followed by fragmentation of the beads. 

Atmospheric moisture and/or water during washing may penetrate into the cracks reached the surface accelerating the corrosion of the beads by glass leaching. 

Changes in glass color around cracks (Fig.\,\ref{fig:destruction}) may be explained by diffusion of impurities. Due to strain stimulated diffusion, like in the gettering process employed in microelectronics, metals may migrate in glass of the green-blue beads to domains of the maximum strain, i.e. to those regions where the cracks arise. 
In addition solid-phase chemical reactions may occur in these domains resulting in formation of some silicates and complex oxides. Glass color may change to green because of accumulation of impurities and formation of the compounds in impurity atmospheres around cracks and bunches of the precipitates.

The observed blackening of glass seen at late phases of corrosion (Fig.\,\ref{fig:destruction}) likely may be explained by purely optical effect conditioned by light reflection on sides of large cracks; light is also scattered by small cracks.\cite{Yuryev_JOPT} Both these phenomena reduce light transmittance as a result of which beads look blackened.
An experiment  on immersion of a blackened bead  in vaseline oil which we carried out supported this statement. After immersion in oil and treating with ultrasound (120 W, 40 kHz), the cracked surface region of the blackened bead brightened. 
In addition, we see that grains of heavily fragmented beads are transparent and have a slightly yellowish or slightly bluish color that also supports the above statement.


It should be emphasized that the processes considered in both scenarios are very long-term.  
They may start at elevated temperature and go on for years and centuries at room temperature due to low rates of all elementary processes involved in the bead corrosion.
Difference in degradation degree of the blue-green bead samples may also be explained by difference in the magnitude of the internal stress and consequently in formation rate of microcracks, diffusion rates of impurities and rates of glass crystallization and solid-phase reactions.
 
If these scenarios of the internal corrosion are true no chemical bead protection techniques will be efficient for conservation of beaded exhibits; the beads will corrode independently of the environmental conditions of their conservation. Physical methods of protection including keeping beaded articles at lowered temperature to decrease rates of impurity diffusion and rates of chemical reactions and processes in the glass matrix and consequently to reduce the rate of the bead deterioration should be developed.

\section{Conclusion}
Summarizing the above, the main results of the article can be formulated as follows.
First, we have discovered antimony-rich crystalline precipitates in the glass bulk of the unstable cloudy blue-green beads and shown them to contain potassium and antimony atoms in equal proportions. Second, we have unambiguously identified the precipitates as orthorhombic KSbOSiO$_4$ crystals.
Concluding the article we would like to emphasize once again that KSbOSiO$_4$ crystallites play the key role in corrosion of  the Sb-rich lead-potassium turquoise  glass  and, as a consequence, in deterioration process of the cloudy blue-green glass beads.

\section*{Supplementary Material}

See supplementary material for EDS spectra obtained at different points on samples of the blue-green beads (Fig.\,\ref{fig:SEM}, Table\,\ref{tab:EDS}).

\begin{acknowledgments}

The research was funded by the Russian Science Foundation (grant No.~16-18-10366).
Analyses by X-ray powder diffraction  were funded through the research project of the Ministry of Culture of RF (State Reg. No.~01201458625).
The research was carried out under the Collaboration Agreement between the State Research Institute for Restoration and A.\,M.\,Prokhorov General Physics Institute of RAS
and
under the Collaboration Agreement between the State Research Institute for Restoration and Kurnakov Institute of General and Inorganic Chemistry RAS.

\end{acknowledgments}


%


\clearpage

\squeezetable
\begin{longtable*}{cccccccccccccccccccc} 
\multicolumn{20}{l}{{TABLE I. Elemental composition  determined using EDS at the points of the bead samples indicated in Figs.\,\ref{fig:SEM} and \ref{fig:SEM-1}.}}\\
\hline\hline 
 Point  & \multicolumn{19}{c}{Elemental composition, wt.\,\% }\\ 
 No.  & \multicolumn{19}{l}{\line(1,0){322} }\\ 
 &O&Na&Mg&Al&Si&S&Cl&K&Ca&Mn&Fe&Cu&Zn&As&Sb&Ba&Au&Hg&Pb\\
\hline
%
%
%
&\multicolumn{19}{l}{Intact blue bead, Fig.\,\ref{fig:SEM}\,a,\,b}\\
1&49.5&0&0&0&22.11&0&0&8.68&0&0&0&1.56&0&0&11.2&0&0&0&7.03\\
2&50.8&0&0&0&27.02&0&0&11.15&0&0&0&0&0&0&0&0&0&0&11.04\\
&\multicolumn{19}{l}{Cleft blue bead (on the cleavage surface), Fig.\,\ref{fig:SEM}\,c,\,d}\\
3&40.7&0&0&12.36&15.79&0&0&7.65&0&0&0&1.95&0&0&21.6&0&0&0&0\\
4&41.2&0&0&16.16&27.67&0&0.83&3.82&0&0&0&3.84&0&0&0&0&0&0&6.47\\
5&45.2&0&0&2.8&16.29&0&0&7.98&0&0&0&1.61&0&0&22.4&0&0&0&3.73\\
6&53.5&0&0&2.66&28.59&0&0&2.96&0&0&0&3.28&0&0&0&0&0&0&9.04\\
7&58.6&0&0&5.22&17.67&0&0&1.47&4.79&0&0&3.24&0&0&0&0&0&0&8.95\\
8&60.3&0&0&4.36&25.39&0&0&3.64&2.28&0&0&4.05&0&0&0&0&0&0&0\\
&\multicolumn{19}{l}{Strongly degraded bead with blue and brown-green (heavily cracked) segments, Fig.\,\ref{fig:SEM}\,e,\,f}\\
9&31&0&0&0.52&11.37&0&0&16.44&0&0&0&0&0&0&40.6&0&0&0&0\\
10&39.2&0&0&0&11.93&0&0&13.75&0&0&0&0&0&0&35.1&0&0&0&0\\
11&54.2&0&0.49&1.42&27.86&0&0&7.48&2.09&0&0&2.6&0&0&0&0&0&0&3.87\\
&\multicolumn{19}{l}{Broken blue-green beads (on the cleavage surface), Fig.\,\ref{fig:SEM}\,g to l}\\
12&52.7&0&0&0&18.04&0&0&8.79&0&0&0&0&0&0&20.5&0&0&0&0\\
13&66.6&0&0&0&24.11&0&0&5.06&0&0&0&1.97&0&0&2.23&0&0&0&0\\
14&51.8&0&0&0&21.61&0&0&6.18&0&0&0&1.99&0&0&14.7&0&0&0&3.72\\
15&57.1&0&0&0&27.73&0&0&3.97&0&0&0&3.42&0&0&0&0&0&0&7.82\\
16&55.4&0&0&1.21&22.94&0&0&6.18&0&0&0&0&0&0&14.3&0&0&0&0\\
17&62.2&0&0&2.24&23.44&0&0&6.86&1.86&0&0&3.46&0&0&0&0&0&0&0\\
18&66.6&0&0&2.83&18.96&0&0&4.52&0&0&0&7.12&0&0&0&0&0&0&0\\
19&61.3&0&0&0&16.06&0&0&7.7&0&0&0&1.14&0&0&13.8&0&0&0&0\\
20&58.8&0&0&0&27.21&0&0&6.66&0&0&0&3.26&0&0&0&0&0&0&4.05\\
21&50&0&0&0&18.01&0&0&10.44&0&0&0&0&0&0&21.6&0&0&0&0\\
22&65.8&0&0&0.75&26.02&0&0&4.51&0&0&0&2.95&0&0&0&0&0&0&0\\
23&57.2&0&0&0&25.93&0&0.48&5.61&0.64&0&0&2.9&0&0&2.76&0&0&0&4.47\\
24&60.7&0&0&0&26.89&0&0.44&5.84&0&0&0&2.33&0&0&0&0&0&0&3.79\\
&\multicolumn{19}{l}{Intact yellow bead, Fig.\,\ref{fig:SEM-1}\,a}\\
25&39.4&1.54&0&0&16.94&0&0&0.7&1.16&0.77&0.61&0&0&0&0&0&0&0&38.9\\
%
&\multicolumn{19}{l}{Cleft yellow bead, Fig.\,\ref{fig:SEM-1}\,b,\,c}\\
26&38.4&0&0&0.99&16.97&0&0&0&0.88&0&0&0&0&0&0&0&0&0&42.81\\
27&33&0&0&0&9.96&0&0&0&1.5&0&0&0&0&0&14.3&0&0&0&41.17\\
28&29.5&0&0&0&10.87&0&0&0&1.6&0&0.99&0&0&0&11.9&0&0&0&45.14\\
29&39.8&0&0&0&16.31&0&0&0.59&0.84&0&0&0&0&0&0&0&0&0&42.5\\
30&28.2&0&0&0&8.64&0&0&0&1.94&0&0&0&0&0&17.2&0&0&0&44.08\\
31&27&0&0&0&8.34&4.1&0&0&1.68&0&0&0&0&0&16.4&0&0&0&42.48\\
&\multicolumn{19}{l}{Intact  opaque white bead, Fig.\,\ref{fig:SEM-1}\,d}\\
32&49.2&1.73&0&0&21.79&0&0&4.69&2.45&0&0&0&0&0&0&0&0&0&20.16\\
33&54.4&1.98&0&0&19.18&0&0&3.43&2.34&0&0&0&0&0&0&0&0&0&18.65\\
&\multicolumn{19}{l}{Intact less opaque white bead, Fig.\,\ref{fig:SEM-1}\,e}\\
34&51.6&2.66&0.82&0.45&18.87&0&0.86&3.78&1.6&0&0&0&0&0&0&0&0&0&19.32\\
35&57.2&3.5&0&0&18.49&0&0&3.02&1.59&0&0&0&0.65&0&0&0&0&0&15.57\\
36&58.8&3.37&0&0.62&19.75&0&0&3.2&1.14&0&0&0&0&0&0&0&0&0&13.15\\
37&55.7&2.84&0.7&0.68&18.56&0&0&2.87&1.18&0&0&0&0&2.36&0&0&0&0&15.08\\
&\multicolumn{19}{l}{Cleft faceted red bead, Fig.\,\ref{fig:SEM-1}\,f}\\
38&63.9&0&0&0&31.07&0&0&2.84&2.17&0&0&0&0&0&0&0&0&0&0\\
39&42.5&0&0&0&11.85&6.98&0&1.3&1.02&0&0&0&0&4.66&0&0&0&31.7&0\\
40&51.7&0&0&0&15.91&8.38&0&1.42&0.96&0&0&0&14.03&0&0&7.66&0&0&0\\
41&58.7&0&1.85&2.5&25.78&0&0&2.79&1.75&0&1.92&0&0&0&0&0&0&0&4.68\\
42&50.4&0&0&0&25.63&0&0&2.29&1.73&0&0&0&0&0&1.93&0&13.6&0&4.4\\
\hline\hline\multicolumn{20}{l}{{\label{tab:EDS}}}\\
\end{longtable*}

~~~

\clearpage

\squeezetable
\begin{longtable*}{cccccccccccccc} 
%
\multicolumn{14}{l}{TABLE II. Reflexes observed in the X-ray diffraction 
patterns (Cu\,K$_\alpha$)  from a powder of three grinded intact blue beads 
(Fig.\,\ref{fig:XRD}); (?) means debatable attribution.}\\
\hline\hline
Line & \multicolumn{2}{c}{Phase} &\multicolumn{4}{c}{$2\thetaup$ (deg)} & Line
&\multicolumn{2}{c}{Phase} &\multicolumn{4}{c}{$2\thetaup$ (deg)} \\
No. & \multicolumn{2}{r}{\line(1,0){85}} &\multicolumn{4}{c}{\line(1,0){110}} & No. &\multicolumn{2}{c}{\line(1,0){80}} &\multicolumn{4}{c}{\line(1,0){110}} \\
 &Structural & Crystal plane  &Tabulated$^{\rm a}$& \multicolumn{2}{c}{Tabulated$^{\rm b}$} & Measured & 
 & Structural & Crystal plane  &Tabulated$^{\rm a}$& \multicolumn{2}{c}{Tabulated$^{\rm b}$} & Measured \\
%
&formula &(h\,k\,l)& K$_{\alpha{_1}}$&K$_{\alpha_{1}}$& K$_{\alpha_{2}}$ &~K$_{\alpha_{1,2}}$ &
&formula &(h\,k\,l)&K$_{\alpha{_1}}$&K$_{\alpha_{1}}$& K$_{\alpha_{2}}$ &~K$_{\alpha_{1,2}}$\\
%
\hline
%
1  &\multicolumn{2}{c}{\textit{Unidentified}}  &&&&11.94&
43  &KSbOSiO$_4$&514&51.1374&51.12&51.25&51.13\\
2  &\multicolumn{2}{c}{\textit{Unidentified}}   &&&&12.97&
44  &KSbOSiO$_4$&613&51.4059&51.37&51.51&$^{~\,}$51.39$^{\rm d}$\\
3  &$^{~\,}$KSbOSiO$_4${$^{\rm c}$} &200&13.6104&13.61&13.64&13.58&
45  &KSbOSiO$_4$&621&&51.42&51.55&$^{~\,}$51.39$^{\rm d}$\\
4  &KSbOSiO$_4$&110&15.2828&15.27&15.31&15.28&
46  &KSbOSiO$_4$&431&51.5369&51.51&51.64&$^{~\,}$51.53$^{\rm d}$\\
5  &KSbOSiO$_4$&201&15.9593&15.97&16.01&15.97&
47  &KSbOSiO$_4$&233&51.5369&51.51&51.65&$^{~\,}$51.53$^{\rm d}$\\
6  &KSbOSiO$_4$&011&16.0264&16.02&16.06&16.02&
48  &KSbOSiO$_4$&424&52.7193&52.70&52.84&52.74\\
7  &KSbOSiO$_4$&111&17.4286&17.42&17.46&17.43&
49  &KSbOSiO$_4$&415&&53.41&53.55&$^{~\,}$53.48$^{\rm d}$\\
8  &KSbOSiO$_4$&211&21.0917&21.08&21.13&21.09&
50  &KSbOSiO$_4$&225&53.4448&53.46&53.61&$^{~\,}$53.48$^{\rm d}$\\
9  &KSbOSiO$_4$&202&21.6001&21.60&21.65&21.62&
51  &KSbOSiO$_4$&622&53.7161&53.70&53.84&53.73\\
10  &KSbOSiO$_4$&112&22.7041&22.70&22.76&22.74&
52  &KSbOSiO$_4$&333&54.0657&54.04&54.18&54.08\\
11  &\multicolumn{2}{c}{\textit{Unidentified}}  &&&&23.30&
53  &KSbOSiO$_4$&530&55.3050&55.26&55.41&55.24\\
12  &\multicolumn{2}{c}{\textit{Unidentified}}  &&&&23.51&
54  &KSbOSiO$_4$&800&&56.57&56.72&56.67\\
13  &KSbOSiO$_4$&212&25.6486&25.65&25.71&25.74&
55  &KSbOSiO$_4$&040&&56.83&56.99&56.84\\
14  &KSbOSiO$_4$&311&26.1171&26.10&26.17&26.14&
56  &KSbOSiO$_4$&433&&57.44&57.60&57.67\\
15  &KSbOSiO$_4$&203&       &28.70&28.78&$^{~\,}$28.75$^{\rm d}$&
57  &KSbOSiO$_4$&713&&57.79&57.94&57.89\\
16  &KSbOSiO$_4$&013&28.7214&28.73&28.80&$^{~\,}$28.75$^{\rm d}$&
58  &KSbOSiO$_4$&811&59.2451&59.20&59.36&59.23\\
17  &KSbOSiO$_4$&312&29.9665&29.96&30.03&30.03&
59  &KSbOSiO$_4$&406&&59.44&59.60&$^{~\,}$59.49$^{\rm d}$\\
18  &KSbOSiO$_4$&411&31.9272&31.91&31.99&$^{~\,}$31.97$^{\rm d}$&
60  &KSbOSiO$_4$&241&59.4842&59.44&59.60&$^{~\,}$59.49$^{\rm d}$\\
19  &KSbOSiO$_4$&221&32.0133&31.99&32.07&$^{~\,}$31.97$^{\rm d}$&
61  &KSbOSiO$_4$&026&59.4842&59.50&59.66&$^{~\,}$59.49$^{\rm d}$\\
20  &KSbOSiO$_4$&402&32.2775&32.26&32.35&$^{~\,}$32.37$^{\rm d}$&
62  &KSbOSiO$_4$&631&&61.18&61.35&$^{~\,}$61.38$^{\rm d}$\\
21  &KSbOSiO$_4$&022&32.3816&32.37&32.45&$^{~\,}$32.37$^{\rm d}$&
63  &KSbOSiO$_4$&035&61.2203&61.19&61.36&$^{~\,}$61.38$^{\rm d}$\\
22  &KSbOSiO$_4$&004&33.7449&33.75&33.84&33.80&
64  &KSbOSiO$_4$&226&61.3536&61.35&61.52&$^{~\,}$61.38$^{\rm d}$\\
23  &KSbOSiO$_4$&222&35.2772&35.26&35.35&35.28&
65  &KSbOSiO$_4$&242&61.5636&61.52&61.69&$^{~\,}$61.38$^{\rm d}$\\
24  &KSbOSiO$_4$&313&35.5504&35.54&35.63&35.55&
66  &KSbOSiO$_4$&207&62.8893&62.89&63.06&$^{~\,}$62.94$^{\rm d}$\\
25  &KSbOSiO$_4$&204&36.5334&36.54&36.64&36.65&
67  &KSbOSiO$_4$&017&62.8893&62.90&63.08&$^{~\,}$62.94$^{\rm d}$\\
26  &\multicolumn{2}{c}{\textit{Unidentified}}   &&&&37.04&
68  &KSbOSiO$_4$&235&63.0414&63.01&63.19&$^{~\,}$62.94$^{\rm d}$\\
27  &KSbOSiO$_4$&~~~~~403 (?)&&37.54&37.64&37.60&
69  &\multicolumn{2}{c}{\textit{Unidentified}}   &&&&63.66\\
28  &KSbOSiO$_4$&~~~~~322 (?)&&38.60&38.70&38.60&
70  &\multicolumn{2}{c}{\textit{Unidentified}}   &&&&63.97\\
29  &KSbOSiO$_4$&413&40.1428&40.13&40.23&$^{~\,}$40.20$^{\rm d}$&
71  &\multicolumn{2}{c}{\textit{Unidentified}}   &&&&64.14\\
30  &KSbOSiO$_4$&223&40.2119&40.19&40.29&$^{~\,}$40.20$^{\rm d}$&
72  &KSbOSiO$_4$&813&64.6975&64.67&64.85&$^{~\,}$64.95$^{\rm d}$\\
31  &KSbOSiO$_4$&512&41.0675&41.05&41.12&41.08&
73  &KSbOSiO$_4$&243&64.9423&64.90&65.08&$^{~\,}$64.95$^{\rm d}$\\
32  &KSbOSiO$_4$&130&42.4569&42.42&42.53&42.49&
74  &KSbOSiO$_4$&534&66.3750&66.34&66.53&66.39\\
33  &KSbOSiO$_4$&323&43.2378&43.21&43.33&$^{~\,}$43.36$^{\rm d}$&
75  &KSbOSiO$_4$&633&66.5993&66.56&66.74&66.72\\
34  &KSbOSiO$_4$&131&43.3337&43.31&43.42&$^{~\,}$43.36$^{\rm d}$&
76  &KSbOSiO$_4$&822&66.7299&66.70&66.89&$^{~\,}$66.88$^{\rm d}$\\
35  &KSbOSiO$_4$&404&44.0184&44.01&44.13&$^{~\,}$44.11$^{\rm d}$&
77  &KSbOSiO$_4$&442&66.9308&66.88&67.07&$^{~\,}$66.88$^{\rm d}$\\
36  &KSbOSiO$_4$&024&44.1025&44.09&44.21&$^{~\,}$44.11$^{\rm d}$&
78  &KSbOSiO$_4$&804&67.5475&67.52&67.71&67.56\\
37  &KSbOSiO$_4$&205&44.8933&44.88&45.00&$^{~\,}$45.04$^{\rm d}$&
79  &KSbOSiO$_4$&044&67.7938&67.76&67.95&67.78\\
38  &KSbOSiO$_4$&611&&44.88&45.00&$^{~\,}$45.04$^{\rm d}$&
80  &KSbOSiO$_4$&625&68.2635&68.23&68.43&$^{~\,}$68.44$^{\rm d}$\\
39  &KSbOSiO$_4$&015&44.8933&44.90&45.02&$^{~\,}$45.04$^{\rm d}$&
81  &KSbOSiO$_4$&435&68.3484&68.31&68.50&$^{~\,}$68.44$^{\rm d}$\\
40  &KSbOSiO$_4$&231&45.0644&45.04&45.16&$^{~\,}$45.04$^{\rm d}$&
82  &KSbOSiO$_4$&715&68.6411&68.62&68.82&$^{~\,}$68.44$^{\rm d}$\\
41  &KSbOSiO$_4$&603&49.2807&49.26&49.39&$^{~\,}$49.37$^{\rm d}$&
83  &\multicolumn{2}{c}{\textit{Unidentified}}   &&&&72.00\\
42  &KSbOSiO$_4$&033&49.4539&49.43&49.56&$^{~\,}$49.37$^{\rm d}$&
84  &KSbOSiO$_4$&733&72.1902&72.14&72.35&72.25\\
\hline\hline
\multicolumn{14}{l}{$^{\rm a}$PDF-2 Powder Diffraction Database, Joint Committee on Powder Diffraction Standards\,--\,International Centre for Diffraction Data.}\\
\multicolumn{14}{l}{$^{\rm b}$Ref.\,\onlinecite{COD-KSbOSiO4}; Crystallography Open Database (COD).}\\
\multicolumn{14}{l}{$^{\rm c}$For all lines: orthorhombic crystal system, space group $Pna2_1$\,(33); CAS No.\,132265-14-6; PDF No.\,00-045-0323. } \\
\multicolumn{14}{l}{$^{\rm d}$A wide band centered at this point. } \\
\multicolumn{14}{l}{{\label{tab:diffract}}}
\end{longtable*}

\end{document}